\documentclass[twocolumn, switch]{article} 

\usepackage{preprint}

\usepackage{amsmath, amsthm, amssymb, amsfonts}

\usepackage{natbib}
\usepackage[utf8]{inputenc}	
\usepackage[T1]{fontenc}	
\usepackage{xcolor}		
\usepackage[colorlinks = true,
            linkcolor = black,
            urlcolor  = black,
            citecolor = blue,
            anchorcolor = black]{hyperref}	
\usepackage{booktabs} 		
\usepackage{nicefrac}		
\usepackage{microtype}		
\usepackage{lineno}		
\usepackage{float}			

\usepackage{lipsum}		
\usepackage{graphicx}

\usepackage{newfloat}
\DeclareFloatingEnvironment[name={Supplementary Figure}]{suppfigure}
\usepackage{sidecap}
\sidecaptionvpos{figure}{c}

\usepackage{titlesec}
\titlespacing\section{0pt}{12pt plus 3pt minus 3pt}{1pt plus 1pt minus 1pt}
\titlespacing\subsection{0pt}{10pt plus 3pt minus 3pt}{1pt plus 1pt minus 1pt}
\titlespacing\subsubsection{0pt}{8pt plus 3pt minus 3pt}{1pt plus 1pt minus 1pt}

\usepackage{tikz,xcolor,hyperref}

\definecolor{lime}{HTML}{A6CE39}
\DeclareRobustCommand{\orcidicon}{
	\begin{tikzpicture}
	\draw[lime, fill=lime] (0,0) 
	circle [radius=0.16] 
	node[white] {{\fontfamily{qag}\selectfont \tiny ID}};
	\draw[white, fill=white] (-0.0625,0.095) 
	circle [radius=0.007];
	\end{tikzpicture}
	\hspace{-2mm}
}
\foreach \x in {A, ..., Z}{\expandafter\xdef\csname orcid\x\endcsname{\noexpand\href{https://orcid.org/\csname orcidauthor\x\endcsname}
			{\noexpand\orcidicon}}
}

\title{\textbf{Of Aliens and Exoplanets \\
Why the search for life, probably, requires the search for water}}

\usepackage{xwatermark}
\newwatermark[firstpage,color=gray!90,angle=0,scale=0.28, xpos=0in,ypos=-5in]{*correspondence: \texttt{darius.modirrousta@inaf.it}}

\usepackage{authblk}

\author[1,2]{Darius Modirrousta-Galian\orcidA{}}
\author[3]{Giovanni Maddalena\orcidB{}}

\affil[1]{INAF – Osservatorio Astronomico di Palermo, Piazza del Parlamento 1, I-90134 Palermo, Italy}
\affil[2]{University of Palermo, Department of Physics and Chemistry, Via Archirafi 36, Palermo, Italy}
\affil[3]{Institute of Quantitative Biology, Biochemistry and Biotechnology/Centre for Science at Extreme Conditions, School of Biological Sciences, University of Edinburgh, Edinburgh, EH9 3FF, UK}

\begin{document}

\twocolumn[ 
  \begin{@twocolumnfalse} 
  
\maketitle

\begin{abstract}
It is not currently possible to create a living organism \textit{ab initio} due to the overwhelming complexity of biological systems. In fact, the origin of life mechanism, this being how biological organisms form from non-living matter, is unknown. In an attempt to better understand how abiogenesis can occur, some researchers have taken water out of their models and instead opted for more exotic approaches. These assumptions will have strong implications for astronomical observations and potential future space exploration. By breaking down water’s properties to the physical, chemical and biological level, herewith it is demonstrated to be the most adequate medium for the formation of life.
\end{abstract}

{\centering \keywords{Astrobiology \and Exobiology \and Abiogenesis \and Water \and SETI}}
\vspace{0.35cm}

  \end{@twocolumnfalse} 
] 



\section{Introduction}

In 1995, the first exoplanet orbiting a Sun-like star, 51 Pegasi b (also known as \textit{Dimidium}), was officially discovered by Michel Mayor and Didier Queloz \citep{Mayor1995}. Over two decades later, in 2019, they earned a Nobel prize, which was a historic moment for the exoplanetary community. Now, as of April 2021, over 4000 exoplanets have been discovered, which is expected to grow as our astronomical instruments improve and our methodologies become more refined. Given the large number of newly discovered worlds, many have considered the possibility of extraterrestrial life. This inquiry is fundamentally important because of its implications for upcoming astronomical missions such as ARIEL and JWST as well as future interstellar space exploration. Knowing where extraterrestrial life forms and thrives may coincide with planets that are more favourable to human life, which may be critical for the human race to survive. The late Stephen Hawking who was a theoretical physicist, cosmologist and director of research at the Centre for Theoretical Cosmology at the University of Cambridge, highlighted the importance of space travel and human survival as such \textit{“I don't think the human race will survive the next thousand years, unless we spread into space. There are too many accidents that can befall life on a single planet. But I'm an optimist. We will reach out to the stars”}. Because of the vastness of the universe it is impossible to explore all or most planetary systems to find the optimum one for future human exploration, so it is essential to have a solid theoretical foundation that can be used to inform our decisions. For instance, knowing when to launch a spacecraft so that it can reach its destination at an optimum time is necessary. Sending it too early means that future technological advances may produce vehicles that would overtake it centuries or millennia after it had been launched. Sending it too late means that a previously sent spacecraft, perhaps being less technologically advanced, would have reached the destination earlier. This conundrum is avoided by performing a theoretical ‘wait calculation’ that considers several parameters such as the growth of technology and the evolution of Earth’s economy \citep{Kennedy2006}. However, if the destination has been erroneously selected then all of the previous planning would be rendered useless. When it comes to interstellar space travel in search of habitable or inhabited planets, having a robust understanding of how biological systems form and thrive is vital for the efficacy of the mission. 

Incidentally, there are several papers focusing on the search for extraterrestrial life that claim, either directly or indirectly, that life (including Earth’s) forms in the absence of water. This argument is incompatible with most lines of evidence and data. Liquid water is essential to life and should be the direction to follow in future space flight missions to improve the likelihood of finding extraterrestrial life. In order to tackle this argument it is appropriate to begin at the cosmochemical level as life is made of matter, so the provenance and abundance of water is relevant for understanding the formation of life. Before progressing, it is essential to issue the caveat that only the usefulness of liquid water, as opposed to vapour or solid ice, is explored. This position is taken as gases are relatively light, so they may not be efficient at mixing different constituents (i.e. weak buoyancy), and solids are typically too viscous for efficient convective mixing within a reasonably short timescale.

\section{Cosmochemical level}

13.8 billion years ago the Universe began from a very hot state that gradually cooled down. Soon after the Big Bang, the Universe was dominated by hydrogen and helium with trace amounts of lithium and beryllium. Gravitational instabilities caused a small fraction of the total hydrogen to form stars. Nuclear fusion within stars led to the eventual chemical enrichment of the Universe. Although there are varying chemical compositions amongst stars, a common practice in astronomy is to adopt our Sun as a reference point. Through astronomical observations of our Sun’s photosphere and laboratory experiments on meteorites, our solar system’s elemental abundances have been determined \citep{Lodders2010},

\begin{table}[h]
\centering
\begin{tabular}{ccc}
\hline
\hline
Element & Number Fraction & Mass Fraction \\
\hline
H       & 0.92            & 0.71          \\
He      & 0.08            & 0.27          \\
O       & $5.4 \times 10^{-4}$        & $8.6 \times 10^{-3}$      \\
C       & $2.5 \times 10^{-4}$        & $3 \times 10^{-3}$        \\
Ne      & $1.1 \times 10^{-4}$        & $2.2 \times 10^{-3}$      \\
N       & $7.2 \times 10^{-5}$        & $10^{-3}$          \\
Mg      & $3.5 \times 10^{-5}$        & $9 \times 10^{-4}$        \\
Si      & $3.4 \times 10^{-5}$        & $9 \times 10^{-4}$        \\
Fe      & $2.9 \times 10^{-5}$        & $1.6 \times 10^{-3}$      \\
S       & $1.5 \times 10^{-5}$        & $5 \times 10^{-4}$        \\
Ar      & $3 \times 10^{-6}$          & $10^{-4}$          \\
Al      & $3 \times 10^{-6}$          & $10^{-4}$          \\
Ca      & $2 \times 10^{-6}$          & $10^{-4}$         
\end{tabular}
\caption{The cosmochemical composition of our solar-system according to \citet{Lodders2010}.}
\label{tab:cosmochemistry}
\end{table}

From Table~\ref{tab:cosmochemistry} it becomes immediately clear how common water is in the Universe as the two elemental constituents of water, hydrogen and oxygen, are the first and third most common elements. It follows that from a purely statistical perspective a strong argument can be made that if life were to exist elsewhere in the universe, it would have most probably interacted with water in a certain manner. However, a large elemental abundance does not imply biological utility. For example, despite being the second most common element in the universe, helium is not largely present within biological systems as it is chemically inert. In order to explore the usefulness of water, one must examine its properties on a physical, chemical and biological level.

\section{Planetary level}

For a planet of a given composition to form, its constituents have to be in a condensed (i.e. icy) state. Therefore, for a water-rich planet to form, it has to be located in a region within the protoplanetary disc where water vapour can form ice (there is no liquid water in the near-vacuum of space, it is thermodynamically forbidden). From fundamental thermodynamic principles, it is known that water vapour can only condense when its vapour pressure is less than its partial pressure, $P_{v}(T) < fP$, where $P_{v}(T)$ is the vapour pressure at temperature $T$, $f$ is the number fraction of water, and $P$ is the total gas pressure. In our solar system the condensation temperature of $\rm H_{2}O$ is $\rm \sim 150~K$ \citep{Podolak2004,Dangelo2015}, which corresponds to a distance of approximately $\rm \sim 3~AU$ from the Sun \citep{Martin2012}. This location is called the ‘frost line’ and its properties will vary depending on the nature of the protoplanetary disc from which the planet forms as well as the spectral type of the host star. Notwithstanding, a value of $\rm \sim 150~K$ is a reasonable benchmark to adopt for purely illustrative purposes. Therefore, for planets that formed at temperatures below $\rm \sim 150~K$, there is a very high probability that water is present in large amounts, such as Uranus and Neptune that are rich in water ice. In contrast, planets that formed at hotter temperatures are unlikely to have a significant abundance of water unless it came from external sources. Earth, for example, is believed to have received most of its water from comets and asteroids \citep{Shiraishi2008}. Whilst the precise mechanism is not fully understood, it is generally accepted to be due to the period of late heavy bombardment \citep{Matter2009}. Even Mars, with its weak gravitational field, shows very strong evidence that it once hosted a substantial amount of water. For instance, high-resolution data from the Mars Reconnaissance Orbiter Context Camera has found topographical evidence for a paleo-fluvial channel system in the Arabia Terra Martian region \citep{Davis2016}. This is indicative of a “warm and wet” Noachian climate model for Mars where abundant precipitation was common. This is further supported by the presence of water-rich materials, such as clays and sulfates on the Martian surface \citep{Sautter2015}. Another example worth mentioning is the discovery of water in the atmosphere of the exoplanet K2-18b \citep{Tsiaras2019,Benneke2019}. This is a super-Earth with a mass and radius of $\rm \sim 8.6M_{\oplus}$ and $\rm \sim 2.6R_{\oplus}$, respectively, that has a temperature well above the condensation point of water \citep[$\rm \sim 265~K$,][]{Benneke2019}. Water has also been found on other warm or hot small exoplanets such as HD~106315~c, HD~3167~c \citep{Guilluy2021}, and LHS~1140~b \citep{Edwards2021}. This is strong evidence that the delivery of water by comets and asteroids could be common not just in our solar system, but elsewhere in the universe. In other words, due to the copious cosmic presence of water, whether a planet formed within or outside of the frost line appears to have little influence on whether it has come in contact with water.

However, to get a more holistic understanding of the presence of water on planetary surfaces, its thermodynamic stability must also be considered. Referring back to Mars, although there is strong evidence supporting the presence of liquid water in its past, it currently has an arid surface. This is believed to be due to solar winds, X-rays and ultraviolet irradiation that stripped away the Martian atmosphere. Therefore, a planet may lose its surface water due to atmospheric erosion. This effect is especially strong if the planet in question has a small mass like Mars. Under those circumstances, life may adapt to its new arid environment through various means such as evolving to thrive deep in the regolith where water may be stored in underground reservoirs. This would result in planets that harbour life but have no surface or atmospherically detectable water. Unfortunately, if the atmosphere has been lost, then there is no known way of detecting potential biosignatures, thus rendering atmospheric spectroscopy obsolete.

Planets are also susceptible to atmospheric evaporation if they orbit too closely to their host star \citep{Zahnle1986,Hunten1987}. Water is not stable when exposed to high temperatures and stellar X-ray and ultraviolet irradiation, thus resulting in its eventual removal \citep{Kurosaki2014,Johnstone2020,Kimura2020}. This is why water is scarce on Venus. The problem with this scenario is that life is very sensitive to high temperatures and it may therefore not survive. For example, to get water to efficiently vaporise so that it remains stably in the atmosphere (i.e. no surface condensation), one requires surface temperatures $\rm \gtrsim 100^{\circ}C$ (this will vary depending on the ambient pressure). At these temperatures only some extremophiles would survive. However, water is a greenhouse gas so its presence in the atmosphere would increase surface temperatures potentially leading to a runaway greenhouse. In other words, in order to get water into the atmosphere so that it is more easily destroyed by stellar irradiation, one risks increasing surface temperatures to lethal levels. In a similar manner, it can be argued that life may adapt by living underground where the temperatures are more favourable but this would result in the same testability problem.

\section{Micropalaeontological level}

The hypothesis that abiogenesis typically occurs under non-aqueous conditions is certainly contradicted by the story of life on Earth. This is incompatible with the fact that the oldest known microfossils are from hydrothermal vents in the Nuvvuagittuq belt in Quebec, Canada. These are filamentous, fossilised microorganisms that are at least 3.77 billion years old but could possibly be as old as 4.28 billion years \citep{Dodd2017}. Conversely, the oldest verified life on land is from 3.5 billion year old stromatolites found in the Dresser Formation, Pilbara Craton, in Western Australia \citep{Djokic2017,Baumgartner2019}. Furthermore, biases in the data must also be considered. For instance, the proportion of Earth’s ‘dry’ surface that has been investigated is much greater than the equivalent for Earth’s oceans. Even with this strong bias, the oldest known microfossils have still been found in the oceans. In addition, considering that most of Earth’s surface is covered by water (approximately $70\%$), it must be acknowledged that there is a strong statistical argument for life’s oldest fossils being located underwater. Moreover, there is a rich body of scientific literature that discusses the many biologically advantageous properties of hydrothermal vents that may have resulted in autocatalytic chemical reactions that contributed to the formation of the first primitive living organisms \citep{Wachtershauser1990,Hordijk2010}. This evidence is dismissed by some with arguments such as the fact that researchers have been unable to artificially form life in hydrous conditions. This is a misleading argument as scientists have never been able to form life \textit{ab initio} under any conditions. However, the field of synthetic biology and more specifically its subfield xenobiology, are bringing us closer to this goal.

\section{Earth is the average planet}

\begin{figure}
\includegraphics[width=\columnwidth]{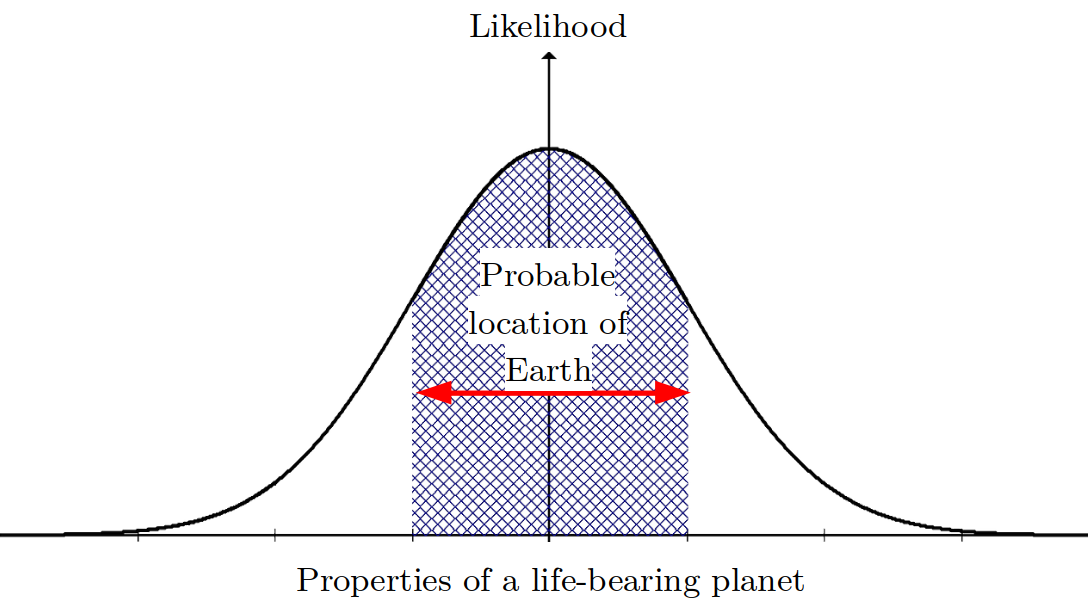}
\caption{If we plotted the properties of life-bearing planets on a graph, Earth would most probably be located somewhere in the centre. Note, this figure assumes that such properties would be distributed in a Gaussian-like manner, the true distribution is unknown.}
\label{fig:Earth_average} 
\end{figure}

The Rare Earth Hypothesis postulates that life on Earth is unique and therefore not a good representation of life elsewhere in the universe. However, a more statistically robust argument would be to assume that Earth’s position is ‘average’ amongst other life-bearing planets (i.e. the \textit{Copernican principle}, see Fig.~\ref{fig:Earth_average}). Astrophysicist, planetary scientist, author, and science communicator, Neil deGrasse Tyson, who is the Frederick P. Rose Director of the Hayden Planetarium at the Rose Center for Earth and Space in New York City, phrased this issue as such: \textit{“Had we been made of some rare isotope of bismuth, you would have an argument to say [that] we are something special”}. Precisely as deGrasse Tyson explained, there is little evidence that supports the argument that biology on Earth is unique within the cosmos. The example given was how the four most common non-inert elements in the universe (hydrogen, oxygen, carbon and nitrogen) are also the most common elements within biology on Earth. These elements, that have the mnemonic acronym of CHON, are accepted by most biological scientists to be the building blocks of life. This is why it is common for many within the petrological community to search for these elements when trying to detect microfossils in billion-year-old rocks when using, for instance, Raman spectroscopy.

\section{Physicochemical level}

To better understand the importance of water in finding extraterrestrial life, the following definition, adapted from commonly used principles, will be applied -- \textit{‘an enclosed membrane which produces energy, and transfers information’} \citep{Dzieciol2012}. Present day terrestrial biology consists of a membrane made of phospholipids and energy in the form of ATP. Within this framework, information is transferred by DNA getting transcribed to RNA, which is then translated to proteins. This is the central dogma of molecular biology \citep{Crick1970}. However, it is worth considering that this optimised process includes billions, or in the very least, hundreds of millions of years of evolution and selective pressures. 

Interestingly, water may have acted indirectly as a catalyst for the evolution of prebiotic molecules, converse to The Water Paradox, which states water breaks down biopolymers (such as proteins and DNA). Mounting evidence suggests that prebiotic chemistry may have originated from wet-dry cycles in shallow sources of water \citep{Frenkel2019}. In the dry phase, water-forming condensation reactions would produce chains between biomolecules, like amino acids and bases for proteins or RNA respectively, with the water then quickly evaporating away. In the wet phase, water returns to the system and breaks down any ‘weak’ links in the biopolymer chains. Regular cycling applies a selection pressure that facilitates the evolution of stable complexes \citep{Frenkel2020}. These biomolecules would interact with each other, or with themselves, to create a steric hindrance, which restricts water’s access to easily hydrolysed chemical bonds.

\begin{figure}
\includegraphics[width=\columnwidth]{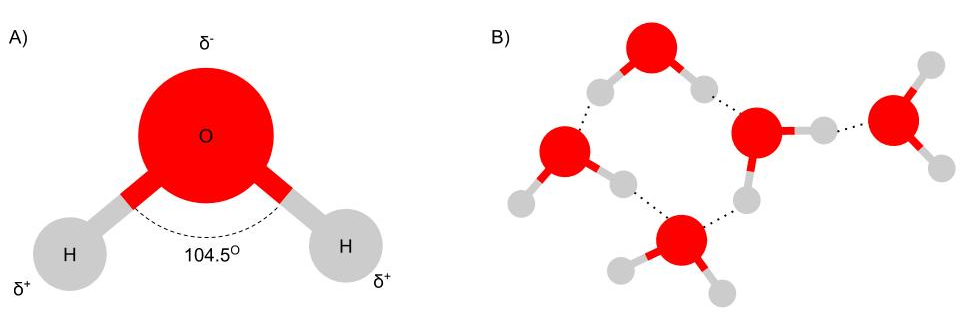}
\caption{The physicochemical properties of water can be explained by the high electronegativity of oxygen, and the low electronegativity of hydrogen, which give it two dipoles capable of hydrogen bonding. The polar nature of the molecule is due to the $104.5^{\circ}$ angle which forms.}
\label{fig:Molecule} 
\end{figure}

The ‘wet-dry’ hypothesis suggests that water was a catalyst for the formation of primordial cells in the early-Earth due to the prevalence of hydrolysis-condensation reactions found in nature. However, this description is unable to fully encapsulate the unique physicochemical properties of water, which may be crucial for abiogenesis. A more holistic analysis can be attained by considering the electronic properties of a $\rm H_{2}O$ molecule. Water consists of two hydrogens and one oxygen. Oxygen is strongly electronegative, second most to fluorine (3.440 and 3.980 respectively), meaning that it has a tendency to attract electrons away from partnered hydrogens (Fig.~\ref{fig:Molecule}A). The H-O-H bond angle is $104.5^{\circ}$ which, when combined with the high disparity in electronegativity, generates a polar molecule with a negative charge density at the oxygen and a positive charge density at the hydrogen. The highly polar nature of water makes it a universal solvent allowing it to dissolve a large range of inorganic and organic molecules. Wet-dry cycles taking place on mineral deposits would benefit from this property as they can solubilise metal ions such as iron and arsenic, which were relatively abundant in the early Earth \citep{Enriquez2012}. The varying redox states of these metals allows them to transfer electrons to other molecules and act as a source of energy \citep{Roden2012,Sforna2014}. Dissolution of these metal ions is an example of how the polar behaviour of water facilitated primordial reactions. The dielectric constant of water is very high (with a value of 81) and it is responsible for weakening the bonds between other molecules such as metal salts in minerals. For comparison, ammonia has a dielectric constant of 16, and hydrocarbons, such as ethane, are nonpolar with a significantly lower constant. The two aforementioned compounds are popular alternatives to water-based life, but their poor ability to solubilise organic and inorganic molecules may prevent this \citep{McKay2014}.

\begin{figure}
\includegraphics[width = \columnwidth]{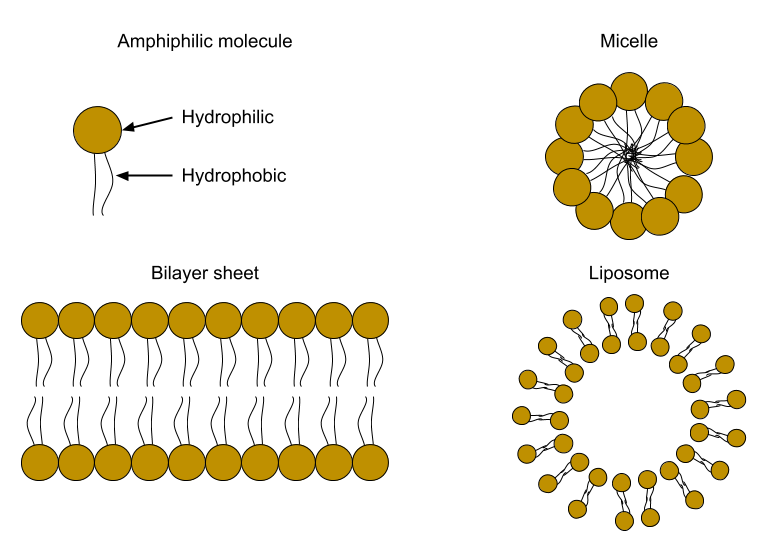}
\caption{The most thermodynamically favourable shapes for amphiphilic molecules to take in water are micelles, bilayer sheets and liposomes. These forms are driven by the hydrophobic effect and may have facilitated the evolution of cellular membranes.}
\label{fig:Folding_molecule} 
\end{figure}

Hydrophobicity is another physicochemical property that is important in sustaining life. This behaviour is crucial to forming an enclosed membrane, a prerequisite to forming enclosed cellular life \citep{Tanford1978,Dzieciol2012}. A semi-permeable barrier provides a means to control the movement of molecules, either produced inside the cell, such as proteins, or carried into the cell, such as metal ions. In aqueous solutions, amphiphilic molecules (containing one charged hydrophilic end and one hydrophobic moiety) will form liposomes, micelles or bilayers as the most thermodynamically favourable structure (Fig.~\ref{fig:Folding_molecule}). This is because these shapes reduce the surface area in contact with water, minimising the free energy of the system.

The high electronegativity of oxygen, in the molecular structure of water, allows it to form hydrogen bonds (Fig.~\ref{fig:Folding_molecule}B). These can form between other water molecules, or with other molecules containing highly electronegative atoms, such as peptides. Hydrogen bonds play a chaperone-like role in biomolecular topography, meaning that they assist in the process of folding biological molecules such as proteins and DNA. The shape of these molecules are key to regulating processes inside and outside of the cell. For example, an enzyme has a very precise and defined shape that dictates the biochemical functions it has. As protein folding takes place, water forms hydrogen bonds with peptides in a rapid cycle of breaking and reforming, following a folding-funnel mechanism. This mechanism causes the potential and configurational energy of a protein to decrease as its native state is approached, following a funnel-like direction as various folding conformations are trialled \citep{Collet2011}. It is the polar nature of water that allows for this process to take place. Without hydrogen bonding and water as a solvent to chaperone, proteins would have an extensive number of probable folding patterns, leading to imprecise folding or no folding at all. This may result in a complete loss of function.

\section{Conclusion}

The only evidence available for life in the universe is that of planet Earth. This limited data sample will result in an unavoidable ‘Earth-centric’ bias, which may have motivated some researchers to opt for more quixotic theories for the formation and evolution of life. Given the vastness of the universe (perhaps being infinite in size), even the least probable events are bound to happen. Therefore, the formation and evolution of exotic forms of life is possible, given that they do not violate the laws of nature. However, just because an event is possible does not mean that it is probable. When it comes to interstellar space travel, one must be as practical as possible due to technological and financial restrictions. Hence, it may be unwise to reject our one single data point in favour of more speculative models for life as these have practical consequences. Whether most life forms on dry land or in liquid conditions will determine the type of exoplanets that are worth investigating spectroscopically and then, potentially, being visited. This is why it is essential to select one’s assumptions reasonably and base them on as much evidence as possible. Understanding the origin of life on Earth and how it could form on other planets, is certainly not an easy task so we admire and encourage those who investigate this issue. We are however cautioning others not to assume, perhaps erroneously, that life formed without water just because the origin of life mechanism has not been discovered yet. Obstacles are expected, after all, nature is mysterious.



\small
\bibliography{references}


\end{document}